\documentclass[twocolumn,preprintnumbers,amssymb,amsmath,9pt]{revtex4}[10pt]
\usepackage{graphicx}
\usepackage{dcolumn}
\usepackage{bm}
\begin{document} 
\input epsf.tex
\newcommand{\beq}{\begin{eqnarray}}
\newcommand{\eeq}{\end{eqnarray}}
\newcommand{\nn}{\nonumber}
\def\ltap{\ \raise.3ex\hbox{$<$\kern-.75em\lower1ex\hbox{$\sim$}}\ }
\def\gtap{\ \raise.3ex\hbox{$>$\kern-.75em\lower1ex\hbox{$\sim$}}\ }
\def\CO{{\cal O}}
\def\CL{{\cal L}}
\def\CM{{\cal M}}
\def\tr{{\rm\ Tr}}
\def\CO{{\cal O}}
\def\CL{{\cal L}}
\def\CM{{\cal M}}
\def\mpl{M_{\rm Pl}}
\newcommand{\bel}[1]{\be\label{#1}}
\def\al{\alpha}
\def\bt{\beta}
\def\eps{\epsilon}
\def\eg{{\it e.g.}}
\def\ie{{\it i.e.}}
\def\mn{{\mu\nu}}
\newcommand{\rep}[1]{{\bf #1}}
\def\be{\begin{equation}}
\def\ee{\end{equation}}
\def\bea{\begin{eqnarray}}
\def\eea{\end{eqnarray}}
\newcommand{\eref}[1]{(\ref{#1})}
\newcommand{\Eref}[1]{Eq.~(\ref{#1})}
\newcommand{\gsim}{ \mathop{}_{\textstyle \sim}^{\textstyle >} }
\newcommand{\lsim}{ \mathop{}_{\textstyle \sim}^{\textstyle <} }
\newcommand{\vev}[1]{ \left\langle {#1} \right\rangle }
\newcommand{\bra}[1]{ \langle {#1} | }
\newcommand{\ket}[1]{ | {#1} \rangle }
\newcommand{\ev}{{\rm eV}}
\newcommand{\kev}{{\rm keV}}
\newcommand{\Mev}{{\rm MeV}}
\newcommand{\gev}{{\rm GeV}}
\newcommand{\tev}{{\rm TeV}}
\newcommand{\mev}{{\rm MeV}}
\newcommand{\mnu}{\ensuremath{m_\nu}}
\newcommand{\mlr}{\ensuremath{m_{lr}}}
\newcommand{\acc}{\ensuremath{{\cal A}}}
\newcommand{\mav}{MaVaNs}
\newcommand{\disc}[1]{{\bf #1}}

\title{Neutrino Oscillations as a Probe of Dark Energy}
\author{David B. Kaplan}
\email{dbkaplan@phys.washington.edu }
\affiliation{ Institute for Nuclear Theory, University of
  Washington,  \\
 Seattle, WA 98195-1550, USA
}%
\author{Ann E. Nelson}
\email{anelson@phys.washington.edu}
\author{Neal  Weiner}
\email{nealw@phys.washington.edu}
\affiliation{
Department of Physics, Box 1560, University of Washington,\\
           Seattle, WA 98195-1560, USA
}
\date{\today}
\preprint{INT-PUB 03-22, UW/PT-03-34 }
\begin{abstract}
We consider a class of theories in which neutrino masses depend
significantly on environment, as a result of interactions with the
dark sector.  Such theories of mass varying
neutrinos (MaVaNs)
were recently introduced to explain the origin of the cosmological
dark energy density and why its magnitude is
apparently coincidental with that of neutrino mass splittings.  In this Letter
we argue that in such 
theories neutrinos can exhibit  different masses in
matter and in vacuum, dramatically affecting neutrino oscillations.
Both long and short baseline experiments are essential to test for
these interactions. As an example of modifications to the standard
picture, we consider simple models which may
simultaneously account for the  LSND anomaly, KamLAND, K2K and studies
of solar and atmospheric neutrinos, while providing motivation to
continue to search for neutrino oscillations in short baseline
experiments such as BooNE.   
\end{abstract}
\maketitle
\section{Introduction and Motivation}
\label{intro}

In the past decade, two of the greatest advances in physics have been the
experimental confirmation of neutrino oscillations, and the observation of acceleration of the
cosmological expansion from a mysterious dark sector. In
this letter, we link the two, discussing how neutrino oscillation experiments
could reveal non-gravitational interactions between matter and the dark sector.

In recent years great progress has been made in understanding neutrino
masses and oscillations.  As first pointed out by Wolfenstein\cite{Wolfenstein:1978ue}, and Mikheev and Smirnov \cite{Mikheev:1985gs,Mikheev:1986wj} the forward
scattering of neutrinos by matter via the weak interactions, while
having a very small cross section, can have a significant effect on neutrino
oscillations.   As a consequence, in  all  theoretical analyses of
the oscillations of neutrinos passing through the sun or earth, matter
effects on neutrino propagation have played a central role.  The
plethora of neutrino mass 
experiments \cite{Smy:2002rz,Wilkes:2002mb,Hallin:2003et,Eguchi:2002dm,Hampel:1998xg,Altmann:2000ft,Abdurashitov:2002nt} appearing in the wake of the original Homestake \cite{Cleveland:1998nv} solar
neutrino experiment have converged to a consistent picture of neutrino
mass: large angle MSW explaining the solar deficit, and a large mixing
angle explaining the atmospheric neutrino deficit as well. 
In spite of this convergence, very little is really known about the
interactions and properties of neutrinos. Aside from mild constraints from BBN
and supernovae, interactions of low energy
neutrinos with themselves or  with ordinary matter are poorly
known.  Given that we already know that neutrino masses have tremendous
environmental dependence even with purely weak interactions, and given our
experimental ignorance of neutrino interactions, we must ask whether new
interactions could offer additional medium dependence, and what physics such new interactions could probe.

A natural origin for new interactions would be the sector responsible
for dark energy. Neutrinos and neutrino oscillations are ideal windows into the
dark sector, not only because the neutrino's known interactions are weak, and
 masses small, but also because 
if lepton number is broken, neutrinos carry no conserved
charges and  are
uniquely capable of mixing with fermions in the dark
sector.  The dark energy offers an important clue, in that its scale - $(2
\times 10^{-3} \ev)^4$ - is comparable to the scale of neutrino mass splittings,
$\delta m_\nu^2 \sim (10^{-2}\, \ev)^2$. Should there be new particles at this
scale, their interactions and mixings with neutrinos could be significant. Given
that neutrino masses are already known to be sensitive to Planck-suppressed
interactions via the seesaw mechanism \cite{seesaw1,seesaw2}, it is not at all
surprising to suggest that new, sub-gravitational forces mediated by the dark
sector particles could generate additional medium dependence of the neutrino
mass.

Here we broaden the discussion of 
 a class of theories first proposed in 
ref.~\cite{Fardon:2003eh} to explain the nature of the dark energy.
These theories explain the similarity between the dark energy scale and the
 measured scale of  neutrino mass splittings,  by postulating 
that neutrino masses are variable, depending on the value of 
a scalar field $\acc$.  The potential for $\acc$ is taken to be very
flat, so that the magnitude of 
$\acc$ depends  upon the cosmological density of neutrinos.  As a
result, these mass varying
neutrinos (MaVaNs) become heavier as their density decreases, and the total energy of the fluid (both in neutrinos and in the $\acc$ field),  identified with the dark energy, can vary slowly as the neutrino density decreases. Not only can this explain the origin of dark energy, but it can also substantially alter the cosmological limits on neutrino mass \cite{Fardon:2003eh}, modify neutrino mass relationships to leptogenesis \cite{Bi:2003yr}, and change the flavor content of the cosmic background neutrinos and distant astrophysical sources \cite{Hung:2003jb}.

 Here we show that sub-gravitational strength interactions
between ordinary matter and the $\acc$ field naturally occur, and can
cause the value of  $\acc$ to differ in the presence of matter from
its vacuum value.  This leads to medium-dependent  neutrino masses and novel 
features in neutrino oscillations.  Observing these effects would not
only extend our understanding of neutrinos, but would also shed light
on the mysterious dark sector which governs the evolution of the
universe on the grandest scales.

After explaining in the next section how such effects arise, we
proceed in   \ref{sec:models} to  consider
how these matter effects can improve agreement between the LSND results and other experiments.

\section{Dark Bosons, Dark Fermions and  the Standard Model}
\label{sec:dark}

For all we know there could be a profusion of new  particles with no
standard model gauge  interactions.  We will refer to such particles
as   ``dark''.  The main constraint on such indiscernible beasts comes
from cosmology---the success of Big Bang nucleosynthesis (BBN)
strongly suggests that the only relativistic species in thermal
equilibrium with visible matter at a temperature of order an MeV are
those we already know about. Thus dark particles must either be much
heavier than an MeV or too weakly interacting  to thermally
equilibrate  with visible matter in the early  universe when the
temperature was a few MeV. Such considerations have been used to place
 limits on neutrino mixing with sterile neutrinos \cite{Barbieri:1990ti,Kainulainen:1990ds}
and on other interactions with light dark particles. Dark particles
are also constrained from the requirement that they not excessively
contribute to supernovae cooling. In ref. \cite{Fardon:2003eh},
however, it was shown that significant neutrino-dark fermion mixing today can be reconciled with BBN and supernova cooling constraints, due to the strong medium dependence of neutrino properties for MaVaNs.

In this section we will explore the potential impact of the dark
sector on  neutrino oscillations.   
We consider a dark  sector consisting of a scalar, $\mathcal A$, and
fermions, $n$. We will take the dark energy scale, $\sim 2\times 10^{-3} \ev$ to be
the typical mass scale of this sector. Taking a cue from the standard
model, where masses range over nearly six orders of magnitude for the
charged fermions, we consider mass parameters within a few orders of
magnitude of the dark energy scale, but not, \eg, Hubble sized Compton
wavelengths as in quintessence models. 

A general Lagrangian for the dark sector includes
\be
\delta {\cal L} = -m_n(\mathcal A) n n-V_0(\mathcal A),
\ee
where we ignore operators involving more than two fermions, which are irrelevant to our
discussion. The Majorana mass $m_n(\mathcal A)$ may be linear in \acc, or some more
complicated function. The only renormalizable interaction allowed
between the dark sector and the standard model  is 
$
y_\nu h \ell n,
$
where $y_\nu$ is the Yukawa coupling of the Higgs boson to a  SM
neutrino and a dark fermion.  
This interaction yields a Dirac mass $m_D=y_\nu v$.  If the dark
fermion Majorana mass is well below the weak scale, $y_\nu$ must be
extremely small. Many simple
mechanisms for  extremely small Yukawa couplings exist in the
literature, for example, see \cite{Dvali:1998qy}. If $m_n(\mathcal A) > m_D$ then
the seesaw mechanism yields an effective $\mathcal A$-dependent
neutrino mass, $m_\nu(\mathcal A) = m_D^2/m_n(\mathcal A)$. We also
assume there may be other contributions to the neutrino mass, \eg\  from
  a GUT-seesaw mechanism, which are \acc\  independent. 

As in \cite{Fardon:2003eh}  the  energy density of the cosmic
background neutrinos will tend to drive $m_\nu$ to smaller values
and, consequently,  $m_n$ to larger values. That is, the effective
neutrino mass is a function of the background neutrino density. The
neutrinos also have an effective coupling to $\mathcal A$ with
strength $\lambda_\nu=\partial m_\nu/\partial \mathcal A\big|_{\vev{A}}$.  
 
For a nonrelativistic neutrino background,  we can find the value of
\acc\ by minimizing the effective potential 
\be
V(\acc) = n_\nu m_\nu(\mathcal A) +  V_0(\mathcal A)\ 
\ee
where $V_0$ is the effective potential in vacuum. 

Up to this point, we have not considered the possible interactions of
$\acc$ with other matter. To begin, we consider  couplings radiatively
generated from SM loops. There are a number of possibilities to consider. The
most potentially significant are corrections to the electron wave function
renormalization (and hence to the electron mass) from W and Higgs loops, and to
the Z-propagator (and hence to quark masses at higher loop). If we consider the
theory to contain just the standard model with variable (\acc-dependent) masses,
these corrections also appear to depend on \acc, at order $G_f \mnu^2$.  
In matter with  density of $3 g/{\rm cm}^3$, such an interaction has a
comparable effect on the \acc\ potential as the cosmic neutrino
background, with the vastly higher density of electrons compensating
for the much weaker coupling. 

However, the electroweak radiative corrections are dominated by high ($\sim M_W$) momenta, so if the $n$ fermions are lighter than $M_W$, they should
also be considered in the loops. A careful treatment finds that the leading
corrections in this case are proportional to $G_f m_D^2$, and independent of \acc. Terms
depending on \acc\ are suppressed by an additional factor of $G_f m_n^2(\acc)$
and are too weak to be relevant. 
We conclude that radiatively generated couplings of the dark scalar to
quarks and charged leptons are not interesting if (and only if) the scalar-neutrino
interaction arises solely due to neutrino mixing with a dark fermion
which is much lighter than the $W$ boson. 

We also consider non-renormalizable operators which couple the dark
scalar to visible matter, such as might arise from quantum gravity. 
At low energies, these interactions would be appear as Yukawa couplings of
\acc\ to the proton, neutron and electron, which we parametrize as
$\lambda_{i} m_{i}/\mpl$, with $i= {\rm p,\, n,\,e}$, respectively, and where $\mpl$ is the Planck
scale. Couplings $\lambda_{n,p} \lesssim 10^{-2}$
are consistent  with  tests of the gravitational  inverse square law
for  an $\acc$ mass larger than $\sim 10^{-11} \ev$ \cite{Adelberger:2003zx}, and, (for $\lambda_p\sim \lambda_n$), with tests of the equivalence principle for an $\acc$ mass larger than $\sim 10^{-8} \ev$ \cite{Su:gu,Smith:cr}.

In the presence of matter, and ignoring the electron contribution,
 we have a new effective potential for \acc
\be
\bar{V} = \frac{ \lambda_B \rho_B \acc}{\mpl} + V(\acc),
\ee
where $\rho_B$ is the mass density of baryonic matter, and we have set $\lambda_p=\lambda_n=\lambda_B$.   

The change in the neutrino mass in the presence of matter may be estimated to be
\be
\Delta
m_\nu\sim 1\,\ev\left(\frac{\lambda_\nu}{10^{-1}}\right)\left(\frac{\lambda_B}{10^{-2}}\right)\left(\frac{\rho_B}{\bar
    \rho_B}\right)\left(\frac{10^{-6} \ev}{m_\acc}\right)^2 
\ee
where $m_\acc^2\equiv V''(\acc)$, and $\bar \rho_B = 3 {\rm g}/{\rm cm^3}$, a baryon mass density which is typical of the earth's crust. 
This estimate assumes the shift in \acc\ is sufficiently small to
allow for a Taylor expansion of $V$ about the present epoch background value for \acc; for a flatter potential, as was
used in \cite{Fardon:2003eh} in order to explain dark energy, the change in the
neutrino mass  will be much larger, and the $\acc$ mass will be variable.

The generic point is that  for MaVaNs, the neutrino mass is  environment
dependent, and the neutrino mass  in rock or in a star can vary
considerably from the neutrino mass in air and in space.  Significant
matter effects on neutrino propagation are familiar, as in the
Standard Model MSW  mechanism. The possibility of such medium dependence was noted early by Wolfenstein \cite{Wolfenstein:1978ue}, and a scenario where Dirac neutrinos only have mass in matter has been considered previously \cite{Sawyer:1998ac}. New scalar contributions to the
effective neutrino mass can be distinguished experimentally  from
standard MSW contributions and  from new vector contributions \cite{Valle:1987gv,Roulet:1991sm,Guzzo:1999iw,Ota:2001pw,Joshipura:2003jh,Davidson:2003ha,Garbutt:2003ih,Campanelli:2003va} as they
are energy independent, and equal for neutrinos and antineutrinos, absent CP violation.

\section{Matter Effects in Existing Experiments}
\label{sec:models}
Since dark energy now provides us a motivation to consider the possibility of medium dependent neutrino mass, we want to examine what the effects could be on existing neutrino data. In particular 
it is instructive to examine the LSND evidence for short baseline $\bar \nu_\mu
\rightarrow \bar \nu_e$ oscillations in light of these
possible matter effects. Here, we will study whether the ability to have different $\delta m^2$'s in air and matter can lead to an improved agreement both between the other positive results as well as the existing negative results.

Let us begin by discussing the relevant experiments. We can loosely group oscillation experiments into three categories. There are long baseline experiments (LBL), which includes solar neutrino experiments, KamLAND, K2K, and SuperK as well as earlier studies of atmospheric neutrinos. These experiments have all seen evidence for neutrino oscillations, and involve significant propagation through dense matter. The positive results are interpreted through neutrino oscillations to require two small mass squared splittings, $\CO(8 \times  10^{-5} \ev^2)$ for the solar neutrinos and KamLAND, and $\CO(2 \times  10^{-3} \ev^2)$ for K2K and atmospheric neutrino studies. The Super-K atmospheric results should not entirely be classified as positive as the through-going muon data show no evidence for oscillation. This result relies on a knowledge of very high (\CO(100 GeV)) neutrinos and may be subject to systematics not present in e.g., the angular dependence of the multi-GeV events. There are null short baseline experiments (NSBL), including the CHOOZ, Bugey, and Palo Verde reactor experiments, and the higher energy CDHS, KARMEN, CHORUS, and NOMAD experiments, involving muon neutrinos. These experiments have all seen no evidence for neutrino oscillations. Lastly, there is LSND, whose results are consistent with oscillations with a mass splitting greater than $3 \times  10^{-2} \ev^2$. These results are summarized in \cite{Hagiwara:fs}.

Because three neutrinos can only accommodate two independent mass splittings, LSND has generally been interpreted to
necessitate an additional sterile neutrino or neutrinos. However,
recent studies (see \cite{Schwetz:2003pv} and references therein) demonstrate that this, too, gives
a poor fit to the data. Incorporating LSND by invoking CPT violation seems in
conflict with recent 
KamLAND data, while a 3+2 sterile scenario \cite{Sorel:2003hf}
improves the fit in a seemingly contrived way by setting the masses of
the sterile neutrinos to lie in regions where the NSBL
constraints are weakest.  

Four neutrino scenarios have a poor fit due in large part to the differences in how neutrino oscillations affect disappearance experiments compared with the positive appearance signal at LSND.
Atmospheric and solar neutrino data are inconsistent with a large
mixing angle  of $\nu_e$ and $\nu_\mu$ with any sterile neutrino, implying that the mass eigenstates associated with solar and atmospheric oscillations are almost entirely active.  Thus in a four neutrino scenario, the  mass eigenstate  associated with the LSND mass squared difference must be mostly sterile, with small admixture of $\nu_e$ and $\nu_\mu$.
With these constraints, the
amplitude for the LSND $\nu_\mu \rightarrow \nu_e$ transition is the
product of two small mixings (the component of the heavy eigenstate which is $\nu_\mu$ and the component which is $\nu_e$), while only one small mixing angle
appears in  the SBL disappearance experiments (the component which is $\nu_e$ for reactor experiments). Put another way, LSND is only sensitive to $\nu_\mu \rightarrow \nu_e$, while disappearance experiments are sensitive to $\nu_e \rightarrow \nu_\mu$ as well as $\nu_e \rightarrow \nu_s$, which is, in general, larger. 

The experimental limits of exotic matter effects on neutrino
oscillations have barely been explored. Here we will see to what extent matter effects can improve agreement of LSND with other experiments.
Of the NSBL experiments, the Bugey experiment involves dominantly propagation through air \cite{Abbes:1995nc,favier}. The Palo Verde results  involve neutrinos dominantly propagating through earth \cite{pcPV}. The CHOOZ experiment neutrinos propagate roughly 10-20\% through earth \cite{pcCHOOZ}. Of the terrestrial positive signal experiments, both KamLAND and Super-K study the propagation of neutrinos through earth.

Within the context of purely three neutrino oscillations, one might want to consider what the limits are on the $\delta m^2$'s and mixing angles in air and earth separately. The possibility that LSND is testing the ``air'' values of the neutrino mass matrix seem excluded by the fact that KARMEN has similar  air pathlength as LSND, and hence would constrain such an oscillation scenario more strongly than ordinary neutrino oscillations.

If one wants to understand the LSND signal from a ``matter'' value for the neutrino mass matrix, there are a number of experiments to consider.
KamLAND gives evidence for large mixing of $\nu_e$ with some other neutrino in earth with a mass splitting $5 \times 10^{-5} \ev^2 \le \delta m^2 \le 10^{-3} \ev^2$ where the upper bound comes from CHOOZ and Palo Verde. Super-K atmospheric and K2K show evidence for $\nu_\mu$ mixing significantly with $\nu_\tau$ and a mass splitting $\delta m^2 \ge 10^{-3} \ev^2$. In fact, the strongest evidence for the scale of the mass splitting comes from the zenith-angle dependence of the multi-GeV events. In this scenario, one has the exciting possibility that the presently quoted value of mass splitting for atmospheric neutrinos is merely an artifact of the significant depletion of those neutrinos originating below the horizon, which could arise in this scenario from a much larger mass splitting in matter. This speculation, however, seems at odds with the through-going muons, which, together with the stopping muons, give an upper bound on the scale of oscillations of about $10^{-2}\ev^2$ \cite{Fukuda:1999pp}.

These results would suggest that using only three neutrinos, one cannot reconcile LSND with the other experiments. However, should there be some systematic effect in the \CO(100 GeV) neutrinos, or if some unknown process contributes to the production of high energy atmospheric neutrinos, one can consider a scenario where $(\nu_\mu+\nu_\tau)/\sqrt{2}$ has a larger mass in matter in order to explain the LSND result, and a mass $\sim 3 \times 10^{-2}\ev$ in air to explain the atmosperhic result. Leaving the lightest two mass eigenstates to be essentially constant in air and matter, and a small admixture of $\nu_e$ in the heaviest, it appears that the matter parameters of  $0.07\, \ev^2 \le \Delta m^2 \le 0.26\, \ev^2$, and $0.02 \lesssim \sin^2 2 \theta \lesssim 0.12$ appears to fit all of the results outside of the throughgoing muons. (The range in the mixing angle could in fact be much larger, depending on the details of how the CHOOZ experiment changes when restricted to limits on matter parameters.)

However, the presence of light SM-singlet states in the theory seems to be necessary for naturalness \cite{Fardon:2003eh}, and so we should also consider the effects on these states in oscillations. Indeed, 
 medium effects  can  improve the fit of four-neutrino scenarios. The medium dependence of the light mass eigenstates arises most simply from changing the mass of the heavy dominantly singlet mass eigenstates. 

The principal limitation on four-neutrino scenarios in the region near $0.1\ \ev^2$ is from Bugey. Since Bugey does not constrain the matter values of the neutrino properties, but only the air, it is straightforward to reconcile LSND with the NSBL experiments. If the singlet state is $\CO(0.3 \, \ev)$ in matter, but in air is much heavier and with smaller mixings, one can achieve a good fit to all existing data.

Of course, by lowering the mass of this singlet state in matter, some dominantly active mass eigenstate should also have a resulting change in its mass. From the LSND result, we expect some mass splitting to change in matter by an amount
\beq
\Delta m^2 > \sin^2 \theta_{\rm LSND}\, \delta m^2_{\rm LSND} \gtrsim 3 \times 10^{-4} \ev^2.
\eeq
This scale suggests that the scenario is very interesting for more precise studies of the differences between air and earth mass parameters, even in existing data sets. A careful study of the implications of the atmospheric neutrino data
would be worthwhile to see whether it is consistent with  different
oscillation lengths in air and in matter. It would be interesting to
see whether  a general fit of the atmospheric  and K2K data  can
constrain four independent parameters 
\be
\delta m_{\rm atm:air}^2, \quad  \delta m_{\rm atm:rock}^2, \quad
\sin^22\theta_{\rm atm:air},\quad    \sin^22\theta_{\rm atm:rock}\ee 
describing $\nu_\mu-\nu_\tau$ oscillations.

A broad set of possibilities exists where the details of the solar neutrinos are affected considerably, and more detailed investigations are underway \cite{uslater}.

\section{Conclusions}
\label{sec:conc}
Although neutrino mass and dark energy are both established elements
of modern physics, the origin of both is unknown. Indeed, many of the
properties and interactions of neutrinos remain undetermined, as are the existence and properties of new dark particles. 
Neutrinos could be significantly affected by interactions with the dark sector  which
are sub-gravitational in strength to other visible matter. Such
interactions can make the neutrino mass a dynamical quantity,
depending on the environment, as well as altering the BBN and
supernovae constraints on neutrino mixing, and cosmological limits on neutrino mass. 

We have demonstrated that sub-gravitational strength matter-neutrino interactions  can  affect our
interpretation of neutrino oscillation experiments, as well as  absolute measurements
of neutrino mass, such as neutrinoless double beta decay and tritium
endpoint experiments.  For instance, such effects could  
improve the agreement of LSND with other experiments. 
Although the mass splittings  which account for  the atmospheric,
solar, K2K and KamLAND experiments apparently indicate
near degeneracy of the three known neutrinos,  this degeneracy may depend on the propagation medium for neutrinos.  Additional short baseline
experiments will provide important tests for MaVaNs, and may directly probe the physics of the cosmological dark energy. Future neutrino experiments should be designed and analyzed with the possibility in mind of matter density dependent neutrino oscillations.

\hskip 0.2in

\noindent {\bf Acknowledgments}

\noindent The authors would like to acknowledge 
Hamish Robertson, Rob Fardon, Wick 
Haxton, Jeff Wilkes, and Kathryn Zurek for useful conversations,  Andr\'e de Gouv\^ea for reading various drafts and giving very useful suggestions, and correspondence and discussions on the attributes of many experiments with 
Bill Louis, Janet Conrad, Giorgio Gratta, Kate Scholberg, Richard Steinberg, Felix 
Boehm, Guido Drexlin, Stuart Freedman, Ed Kearns, Chris Walter and Jean Favier.
This work was partially supported by the DOE under contract DE-FGO3-96-ER40956 and DE-FGO3-00ER41132.

\end{document}